\documentclass[runningheads]{llncs}

\usepackage[T1]{fontenc}
\usepackage{float}
\usepackage{graphicx}
 \usepackage{booktabs}
 \usepackage[hidelinks]{hyperref}
 \usepackage{subcaption}
 \usepackage{caption}
\usepackage{biblatex}
\begin{document}

\title{Global Crises and National Policies: A Large Scale Analysis of Political Content in German Language Online Media}

\titlerunning{Global Crises and National Policies: Analyzing Political Content in Media}

\author{Yara Döring\inst{1} \and
Felix Bießmann\inst{1,2}\orcidID{0000-0002-3422-1026}}
\authorrunning{Döring and Bießmann}
%
\institute{Berliner Hochschule für Technik \and
Einstein Center Digital Future, Berlin
\email{felix.biessmann@bht-berlin.de}}

\maketitle

\begin{abstract}
Today most media content is consumed based on algorithmic recommendations. Evidence suggests that this can lead to politically biased media consumption patterns. Automated extraction of political agendas from texts can reveal and analyze political biases in online media -- and thus help fostering politically unbiased media consumption. Here we employ modern political text analysis methods demonstrating the potential of automated fine-grained political bias analysis in online media. We conduct an analysis of political content in German language online media during the period 2019--2022, encompassing several million articles and tweets covering events with profound societal impact globally and nationally, the COVID-19 pandemic and the beginning of the war in Ukraine. Our analysis identifies thematic similarity between national (German and Swiss) reporting, particularly for categories driven by international events. We also find divergences emerging in domestically influenced categories, reflecting differences in national policies and institutional structures. A comparison of newspaper and Twitter discourse reveals that both media converge around a shared core during the pandemic, yet differ in intensity and temporal dynamics. Newspapers exhibit more stable political content, while Twitter reacts through short-lived event-driven spikes. These findings indicate that international crises act as a powerful synchronizing force on political content in classical media, temporarily overriding both national and media-form differences. Our automated political analysis empowers citizens by rendering political agendas in online media transparent, enabling societies to critically evaluate content, rather than passively absorbing content curated by opaque algorithms. This transparency also encourages media outlets to be more accountable for their reporting and helps to bridge the gap between algorithm-driven echo chambers and a more informed, balanced public discourse.

\keywords{Political text classification \and Natural language processing  \and Manifesto Project \and Media analysis \and COVID-19 \and Comparative media research \and social media}
\end{abstract}

\section{Introduction}

The majority of media content is consumed through algorithmic recommendations. These recommendations are often biased by economic interests of online media platforms. While the effects on users are an active subject of research, users are exposed to information that can reinforce politically biased consumption patterns~\cite{wagner2021affective,piccardiRerankingPartisanAnimosity2025}. The opaque nature of these algorithms raises concerns about the potential for echo chambers and the polarization of public discourse. To address these challenges, the automated extraction and analysis of political agendas from textual data offer a promising approach to uncovering and understanding political biases in online media~\cite{stier2023algorithmically}. By making these biases transparent, such methods can empower citizens to engage more critically with media content and foster a more balanced and informed public sphere.

One of the drawbacks of many prior studies in this field is that the political content is often analyzed in a very coarse grained fashion, typically only a one-dimensional spectrum of two parties or sentiments. In this study, we leverage modern political text analysis techniques to demonstrate the feasibility and value of automated, fine-grained political bias analysis in online media. 

Our research focuses on German-language media during the period from 2019 to 2022, a time marked by globally significant events such as the COVID-19 pandemic and the onset of the war in Ukraine. The COVID-19 pandemic constitutes one of the most significant global health crises in recent history, with hundreds of millions of confirmed infections worldwide~\cite{noauthor_covid-19_nodate}. In the German-speaking DACH region (Germany, Austria, Switzerland), public demand for reliable information from established news sources increased substantially during this period~\cite{dreisiebner_information_2022}. To study the coverage of these events in online media we analyze a comprehensive dataset comprising several million articles and tweets, allowing for an examination of fine-grained political content across both traditional news outlets and social media platforms.

Our findings reveal notable thematic similarities in national reporting between Germany and Switzerland, particularly in categories dominated by international events. This convergence suggests that global crises can act as powerful synchronizing forces, temporarily aligning political narratives across borders. However, we also observe divergences in domestically influenced categories, where differences in national policies, institutional structures, and public priorities become apparent. These variations highlight the interplay between global events and local contexts in shaping media discourse.

A comparative analysis of newspaper and Twitter content further underscores the distinct dynamics of political communication in different media forms. While both platforms are impacted strongly by global crises such as the COVID pandemic, they exhibit marked differences in intensity and temporal patterns. Traditional newspapers tend to maintain a more stable and sustained focus on political issues, whereas Twitter discourse is characterized by short-lived, event-driven spikes in attention. These differences reflect the unique affordances and audience behaviors associated with each medium.

The results of our study underscore the potential of automated political analysis to enhance transparency in media reporting. By rendering political agendas more visible, our approach enables citizens to critically evaluate the content they encounter, rather than passively absorbing information curated by opaque algorithms. This transparency not only empowers individuals to make more informed choices but also encourages media outlets to be more accountable for their reporting practices. Ultimately, our work contributes to bridging the gap between algorithm-driven echo chambers and a more informed, balanced, and participatory public discourse.



\section{Related Work}

Political preferences in online media have been studied in various research communities~\cite{ceron2024large}. In the information retrieval and Machine Learning (ML) communities, modern text analysis methods have increasingly been applied to large-scale media corpora, combining supervised classification frameworks with transformer-based language models. The automated detection of political bias in news media has been approached from multiple angles, including hyperpartisan news detection, identifying articles that take extreme left- or right-wing standpoints. A SemEval shared task demonstrated both the feasibility and challenges of automating such detection, with the best-performing system achieving an accuracy of 0.822 on a manually labeled dataset~\cite{kiesel_semeval-2019_2019}.

These automated extraction methods of political views have also been applied to investigate global crises such as the COVID pandemic. During the COVID-19 pandemic, citizens increasingly turned to established sources such as public broadcasting and national newspapers~\cite{dreisiebner_information_2022}. Cross-cultural comparisons of newspaper coverage revealed significant differences in thematic framing across culturally distant countries~\cite{sato_cross-cultural_2022}, while within the German-language space, German outlets published significantly more articles on solidarity than Swiss ones, a difference linked to stricter COVID-19 measures in Germany~\cite{zimmermann_newspaper_2023}. An active topic of scientific research is the interdependency of classical media and online social media. Certain pandemic policy debates were found to be initially led on social media, with traditional newspapers subsequently following their lead, suggesting a bidirectional relationship between platforms that motivates the comparative analysis conducted here~\cite{gilardi_social_2022}.

One major drawback of most existing studies on political preferences in online media have been severely limited by the spectrum of political views they investigate -- often due to the fact that the political system in some countries, such as the United States, has only two parties. Other democratic nations have parliaments with more than two parties, accounting for and requiring a more fine-grained analysis of political preferences. The Manifesto Project coding scheme allows for such a fine-grained analysis, which organizes political statements into 56 categories spanning seven policy domains and has since been extended beyond party manifestos to other text types~\cite{lehmann_manifesto_2025}. However, models trained on party manifestos are not always directly applicable to other text domains, such as news or online social media. There is work that used the Manifesto Coding scheme for annotating a large German Twitter dataset (2019--2022)~\cite{biessmann_2022_7635612}. This data has been used to fine tune a Transformer model and analyse Tweets during the COVID-19 pandemic, which revealed a marked increase in political expression, particularly in welfare and education~\cite{biessmann2022changes}. 

To summarize while there are studies on automated political text analysis, most of them are limited by their simplistic account of political preferences. Models that account for more fine-grained political analysis are often focusing on one text domain only, such as party manifestos~\cite{burst_manifestoberta_2023} or online social media~\cite{biessmann2022changes}. In order to study the interdependencies between classical and modern social online media, we annotated and used a novel data set of annotated news articles and tweets to fine tune a model for automated finegrained political text analysis of both types of text domains. 

\section{Methods}
In the following we describe the data acquisition process as well as the analysis methods for extracting political preferences from texts. 

\subsection{Data}
News articles were collected using the open-source news crawler Fundus~\cite{noauthor_fundus_nodate} via the CC-NEWS archive, a continuously updated collection of global news content. Data was gathered for the years 2019 to 2022 in German, covering outlets from Germany, Austria, and Switzerland. Due to insufficient Austrian coverage in the CC-NEWS archive, the final dataset is limited to Germany and Switzerland. Crawling was executed as Kubernetes jobs on the university cluster, using 15 CPU cores and 15 GB RAM.
Each article was stored in JSONL format containing the fields \texttt{title}, \texttt{url}, \texttt{plaintext}, \texttt{publishing\_date}, and \texttt{publisher}. After filtering articles outside the study period, the final dataset comprises 3,221,998 articles (see Table~\ref{tab:data}).

\begin{table}[ht]
\centering
\caption{Number of crawled news articles by year and country.}
\label{tab:data}
\begin{tabular}{lrr}
\toprule
\textbf{Year} & \textbf{Germany} & \textbf{Switzerland} \\
\midrule
2019  & 489,162   & 2,824  \\
2020  & 655,012   & 31,705 \\
2021  & 784,272   & 30,675 \\
2022  & 1,195,600 & 32,729 \\
\midrule
\textbf{Total} & \textbf{3,124,046} & \textbf{97,933} \\
\bottomrule
\end{tabular}
\end{table}

\begin{figure}[H]
    \centering
    \includegraphics[width=0.7\textwidth]{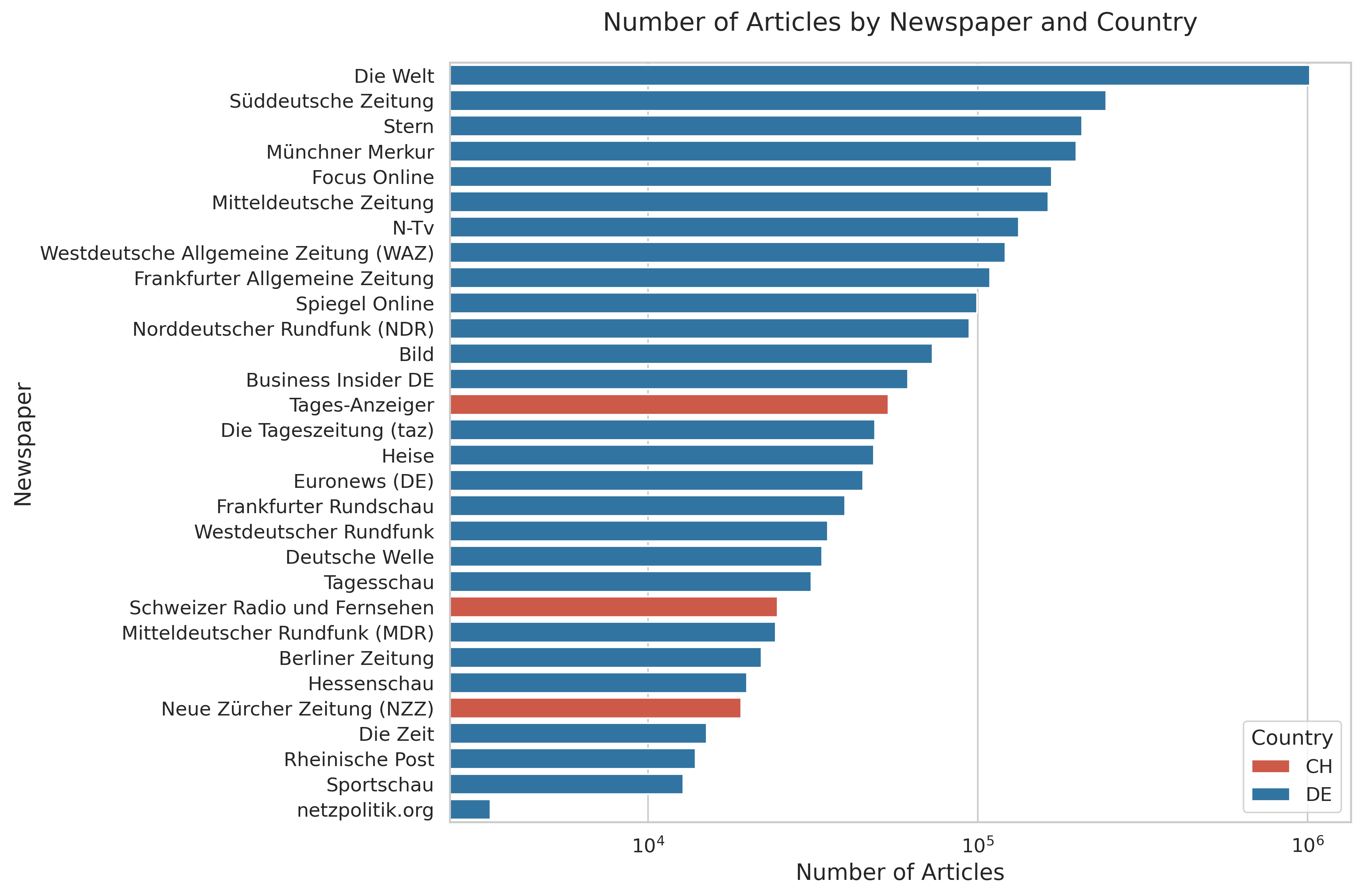}
    \caption{Distribution of articles across newspapers in Germany (DE) and 
    Switzerland (CH). The dataset is dominated by German outlets, with only 
    three Swiss newspapers included: Tages-Anzeiger, Schweizer Radio und 
    Fernsehen, and Neue Zürcher Zeitung (NZZ).}
    \label{fig:publisher_distribution}
\end{figure}
The news data was extended with a dataset of German-language tweets, based on an openly available history Tweet corpus~\cite{kratzke_monthly_2020}. Weekly samples were drawn from the corpus, covering the period 2019-2021 and filtered using a model trained on the Manifesto Codes using annotated Twitter data from the politweets corpus~\cite{biessmann_2022_7635612}, replicating the filtering procedure proposed in~\cite{biessmann2022changes}. The resulting dataset  contains 331,819 tweets of the 56 Manifesto categories.

\subsection{Classification Model and Extraction of Political Views}
%
For automatic extraction of political views we used a model based on an XLM-RoBERTa, finetuned on the 56-category Manifesto Project coding scheme~\cite{lehmann_manifesto_2025}. The pretrained ManifestoBerta model~\cite{burst_manifestoberta_2023} was further finetuned on a data set comprising Tweets and news articles. The Twitter data was obtained from the politweets data set~\cite{biessmann_2022_7635612} and contained 10,561 distinct tweets. The news paper article used for fine tuning were a sample of 1119 articles sampled from the fundus corpus\cite{dallabetta-etal-2024-fundus}. They were manually annotated with the 56 fine-grained political categories~\cite{lehmann_manifesto_2025} by political scientists trained by the Manifesto Coding team. The resulting model used in this study obtained macro averaged recall, precision and F1 scores of 0.60, 0.58 and 0.58, respectively, which can be considered competitive compared to the ManifestoBerta model~\cite{burst_manifestoberta_2023} evaluated on party manifestos and is released on HuggingFace\footnote{\url{https://huggingface.co/rejtrace/politweets-new/}}.
%
Articles exceeding a length of 512 tokens were split into non-overlapping chunks of up to 512 tokens. Each chunk was classified independently, producing a probability distribution over all 56 categories. Chunk-level probabilities were subsequently averaged per article to yield a single article-level distribution. The label \texttt{000-Undefined} was excluded from the analysis as it carries no thematic content. The analysis period was further restricted to December 2019--December 2022 to ensure comparable Swiss data availability.
Classification of articles and tweets was executed on a Kubernetes cluster using an NVIDIA A100 GPU (40 GB) with 16-bit floating-point precision. Processing the full dataset of approximately 5.4 million chunks required approximately 18 hours.

\subsection{Analytical Procedure}
Monthly thematic ratios were computed for each Manifesto category and source (Germany, Switzerland, Twitter) by summing predicted frequencies of articles across all articles in a given month and normalizing so that the 56 category shares sum to 1. Thematic similarity between Germany and Switzerland was assessed via Pearson correlation of these monthly ratios across the shared observation period, retaining only categories with at least three common observations and $p < 0.05$.
For the ideological analysis, categories were mapped to left and right dimensions following the Manifesto Project's RILE scale~\cite{lehmann_manifesto_2025}. Left- and right-coded probability masses were computed per article and aggregated as monthly means.

\section{Results}
In the following we analyse the political content across classical news media and online social media, as well as across German speaking nations (Germany and Switzerland). We focus on similarities in the development of the political discourse and the differences, both with respect to the overall topic distribution but also with respect to the temporal dynamics. 
\subsection{Political Topics in News and Twitter During COVID-19}
The pre-pandemic phase reveals thematic priorities for news outlets and social media, respectively, as illustrated in Fig.~\ref{fig:domain_trends}. The german newspaper agenda is stable and dominated by \textit{Economy} topics (25-30\%), while Twitter shows a more diffuse distribution with no single domain persistently exceeding 20 \%. 
\begin{figure}[H]
    \centering
    \includegraphics[width=\textwidth]{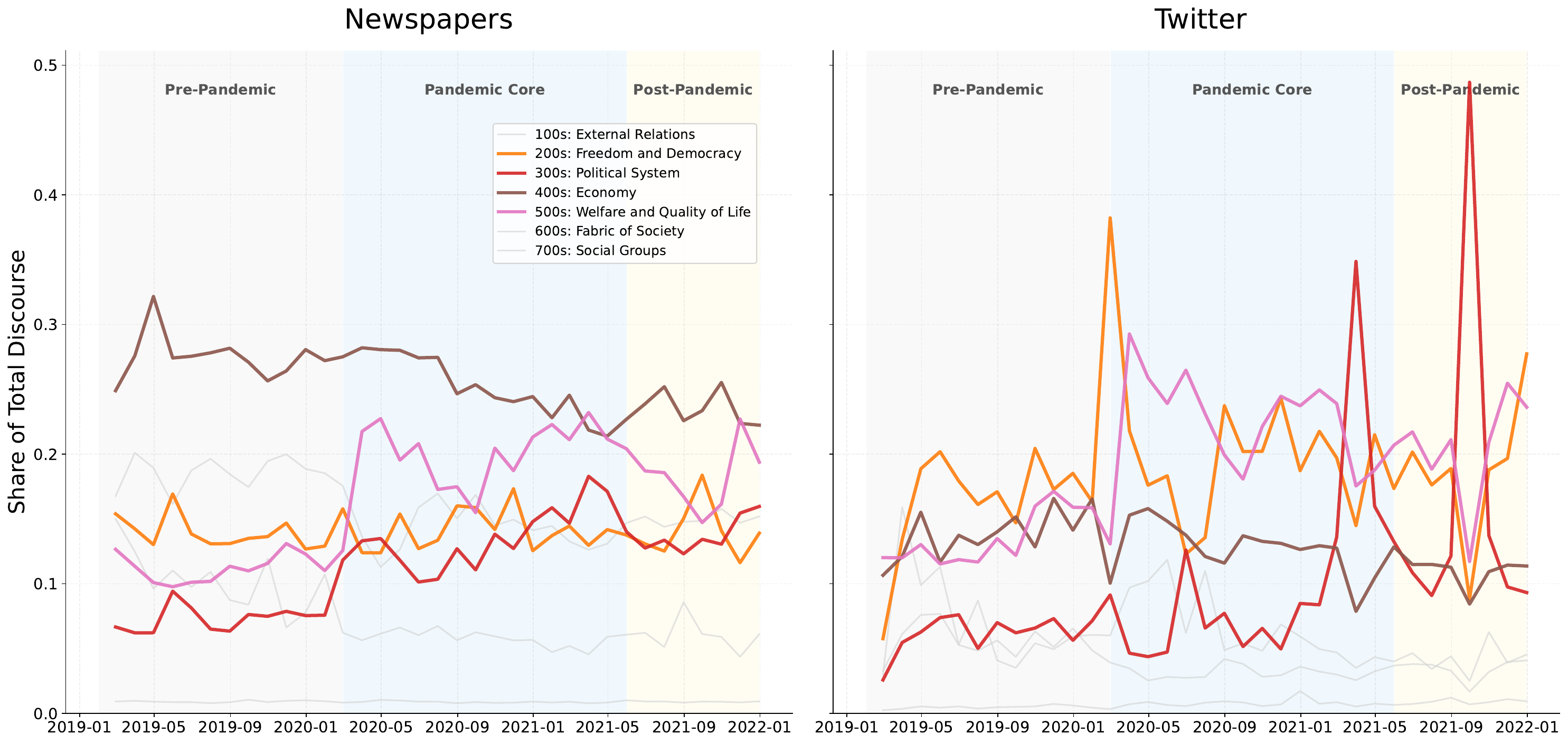} 
    \caption{Longitudinal trends in thematic domain shares across newspapers and Twitter across pre-pandemic, pandemic core, and post-pandemic phases. Newspapers exhibit relatively stable distributions throughout, while Twitter shows considerably higher volatility,  particularly during the pandemic core phase, where several domains spike sharply before subsiding.}
    \label{fig:domain_trends}
\end{figure}

\begin{figure}[H]
    \centering
    \includegraphics[width=\textwidth]{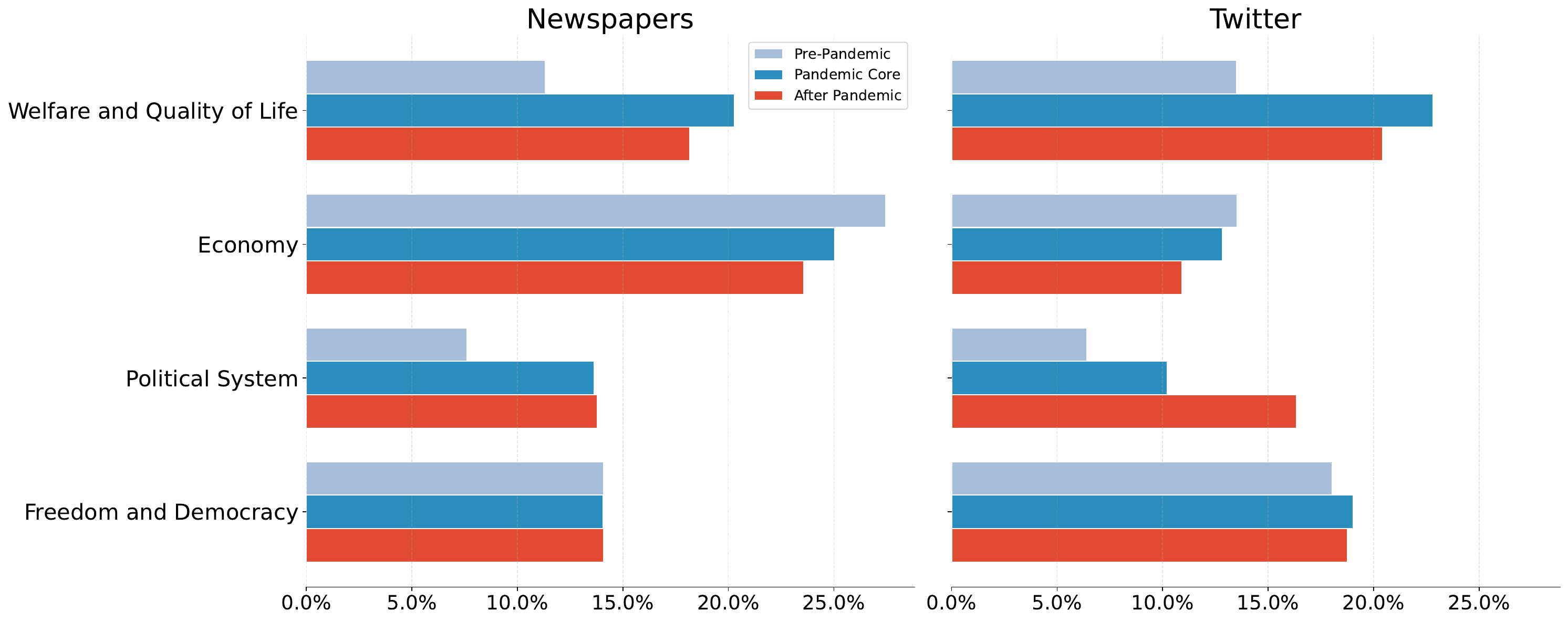} 
    \caption{Distribution of thematic domains in newspapers and Twitter across three pandemic phases (pre-pandemic, pandemic-core, post-pandemic), revealing a convergence in topic coverage between traditional and social media during COVID-19.}
    \label{fig:bar_domain}
\end{figure}
\begin{figure}[H]
    \centering
    \includegraphics[width=\textwidth]{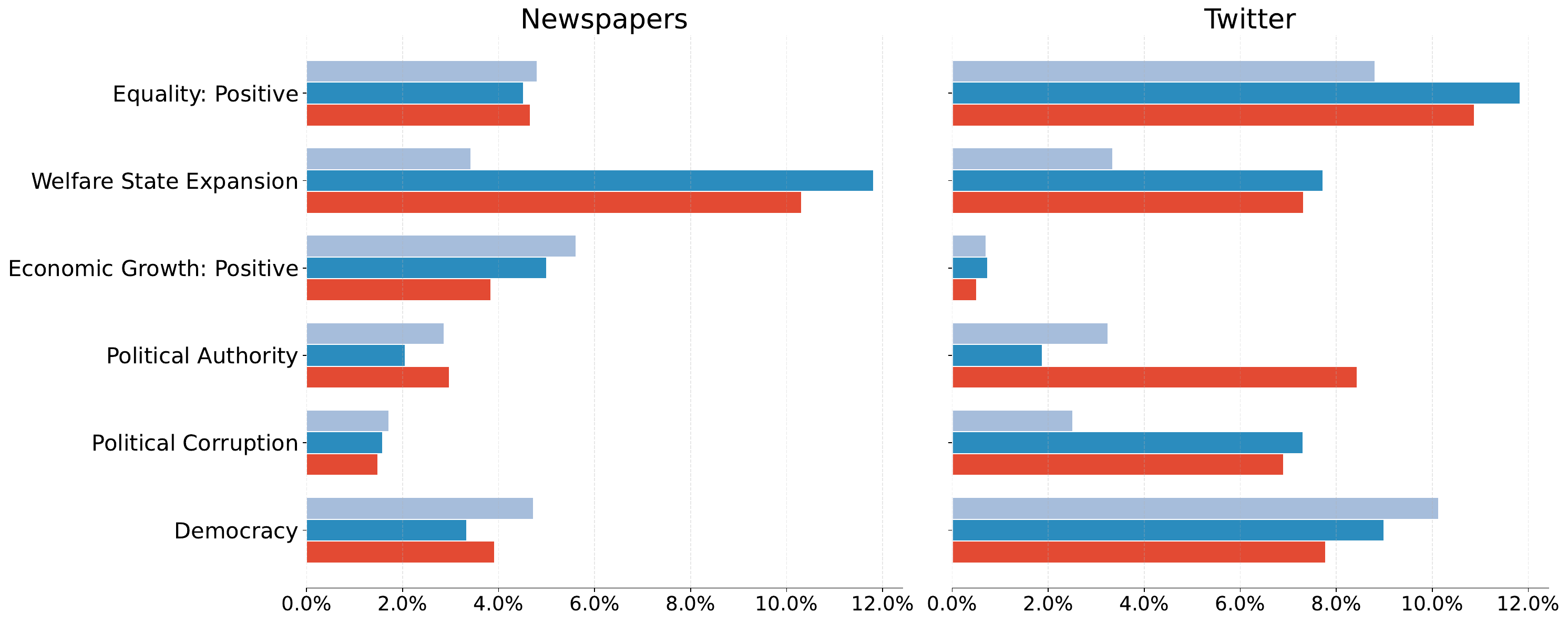} 
    \caption{Distribution of individual thematic labels in newspapers and Twitter across pandemic phases. In contrast to the overarching domain level, individual labels exhibit pronounced differences between the two media, with Twitter showing more volatile and extreme shifts in topic salience.}
    \label{fig:label_trends}
\end{figure}

During the core pandemic phase (2020--2021), both agendas shift towards a shared 
thematic core. The domain \textit{Welfare and Quality of Life} rises in both media at the onset of the pandemic and remains elevated throughout the phase. At the aggregated domain level, the distribution of newspapers and Twitter aligns closely across all three phases, with both media showing comparable ratios for each domain (Figure \ref{fig:bar_domain}). At the level of individual labels, this alignment breaks down (Figure \ref{fig:label_trends}). In the newspaper corpus, \textit{Welfare State Expansion} dominates the core pandemic phase, while the remaining labels stay comparably low. The Twitter corpus, by contrast, shows a much more volatile pattern, with multiple labels rising sharply during the pandemic core phase, including \textit{Equality:Positive} and \textit{Political Corruption}. This volatility is further confirmed by the spikes visible in Fig. \ref{fig:domain_trends}, where \textit{Freedom and Democracy} peaks sharply in March 2020 and \textit{Political System} rises in spring 2021 on Twitter, both collapsing rapidly after. In the newspaper corpus \textit{Political System} also increases during this phase, but without a comparable spike. 

In the period after the core pandemic (end of 2021), the two media forms diverge sharply. On Twitter, the \textit{Freedom and Democracy} domain rises to nearly 45 \% around the federal election in Germany, overshadowing all other domains before collapsing rapidly. In the newspaper corpus, changes remain moderate, with only a small peak in \textit{Political System} in period of the election, while \textit{Welfare and Quality of Life} declines gradually before recovering to its previous level shortly after. Together, these patterns show that while the pandemic temporarily synchronized both media forms, this convergence was not sustained beyond the immediate crisis.

\subsection{Thematic Similarity between German and Swiss News}

Investigating similarities between classical news outlets in Germany and Switzerland we find that the majority of the 56 Manifesto categories show significantly positive correlations between the two German speaking nations. The strongest correlations appear in internationally driven or globally relevant categories: \textit{Military: Negative} ($\rho = 0.93$), \textit{Welfare State Expansion} ($\rho = 0.89$), and \textit{Anti-Imperialism} ($\rho = 0.85$). Notably, five of the nine most correlated categories have a pronounced international character. This suggests that the media coverage and political discourse in both countries react synchronously to global events.
\begin{table}[H]
\centering
\caption{Top and bottom thematic correlations of monthly ratios (DE vs. CH)}
\label{tab:korrelationen_short}
\begin{tabular}{lc}
\toprule
\textbf{Categories} & \textbf{Correlation} \\
\midrule
105 - Military: Negative & 0.93 \\
504 - Welfare State Expansion & 0.89 \\
303 - Governmental and Administrative Efficiency & 0.85 \\
505 - Welfare State Limitation & 0.85 \\
103 - Anti-Imperialism & 0.85 \\
106 - Peace & 0.83 \\
102 - Foreign Special Relationships: Negative & 0.81 \\
\midrule
$\dots$ & $\dots$ \\
\midrule
301 - Federalism & 0.34 \\
415 - Marxist Analysis: Positive & 0.32 \\
603 - Traditional Morality: Positive & 0.32 \\
403 - Market Regulation & 0.28 \\
305 - Political Authority & 0.24 \\
703 - Agriculture and Farmers: Positive & 0.20 \\
302 - Centralisation & 0.10 \\
506 - Education Expansion & 0.09 \\
706 - Non-economic Demographic Groups & -0.18 \\
\bottomrule
\end{tabular}
\vspace{-10pt}

\end{table}

In contrast, categories with the lowest correlations within the dataset are predominantly found in areas of domestic policy and national institutional structures. While some of these correlations remain statistically significant, they are considerably weaker relative to the rest of the dataset. This is particularly evident in categories such as \textit{Centralisation} ($\rho = 0.10$), \textit{Education Expansion} ($\rho = 0.09$), and \textit{Non-economic Demographic Groups} ($\rho = -0.18$), the only category showing a negative correlation. Categories such as \textit{Federalism} ($\rho = 0.34$) and \textit{Political Authority} ($\rho = 0.24$), while positive, rank among the lowest in the dataset. 

\begin{figure}[H]
    \centering
    \begin{subfigure}[b]{0.48\textwidth}
        \centering
        \includegraphics[width=\textwidth]{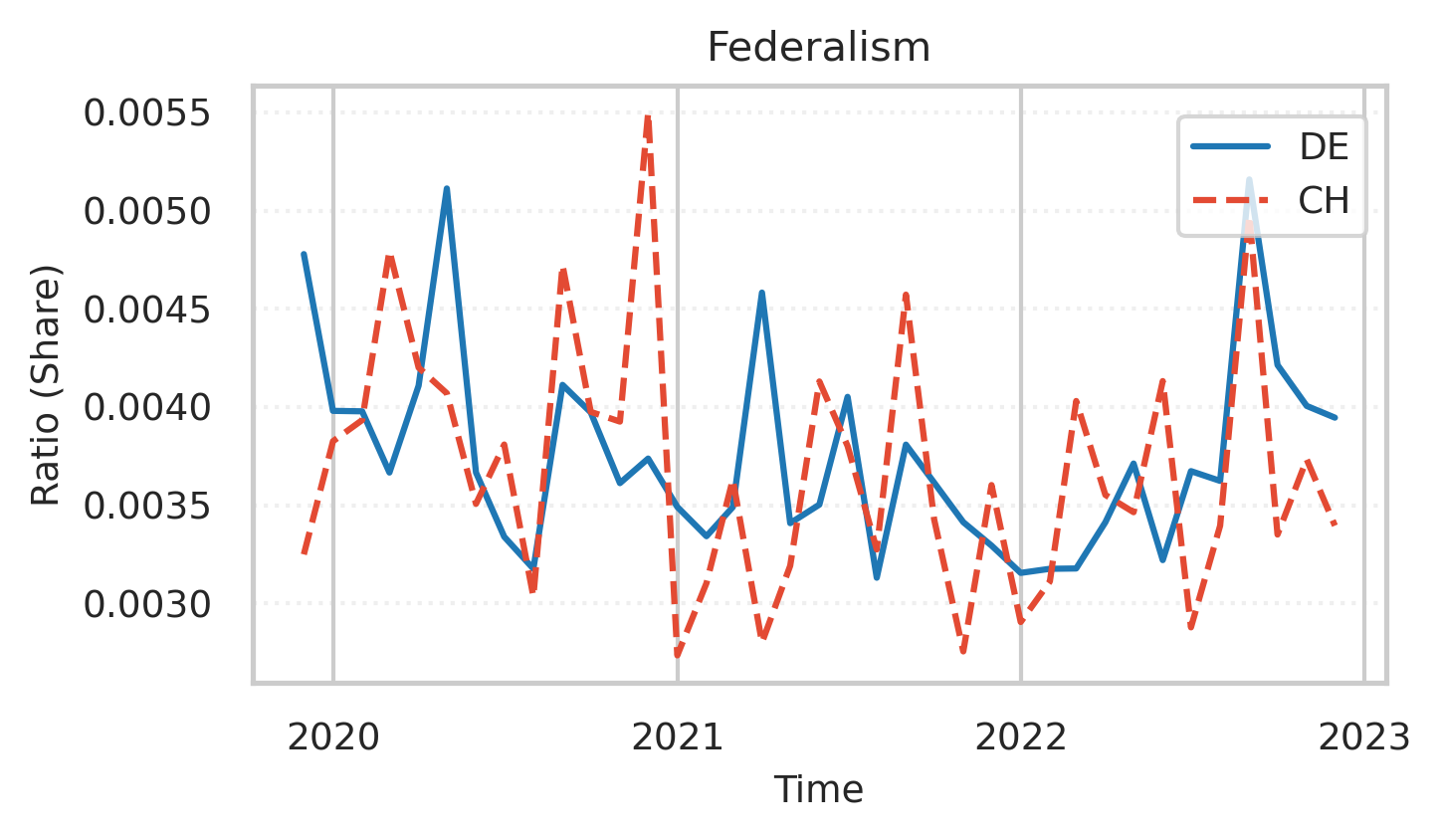}
        \label{fig:federalism}
    \end{subfigure}
    \hfill
    \begin{subfigure}[b]{0.48\textwidth}
        \centering
        \includegraphics[width=\textwidth]{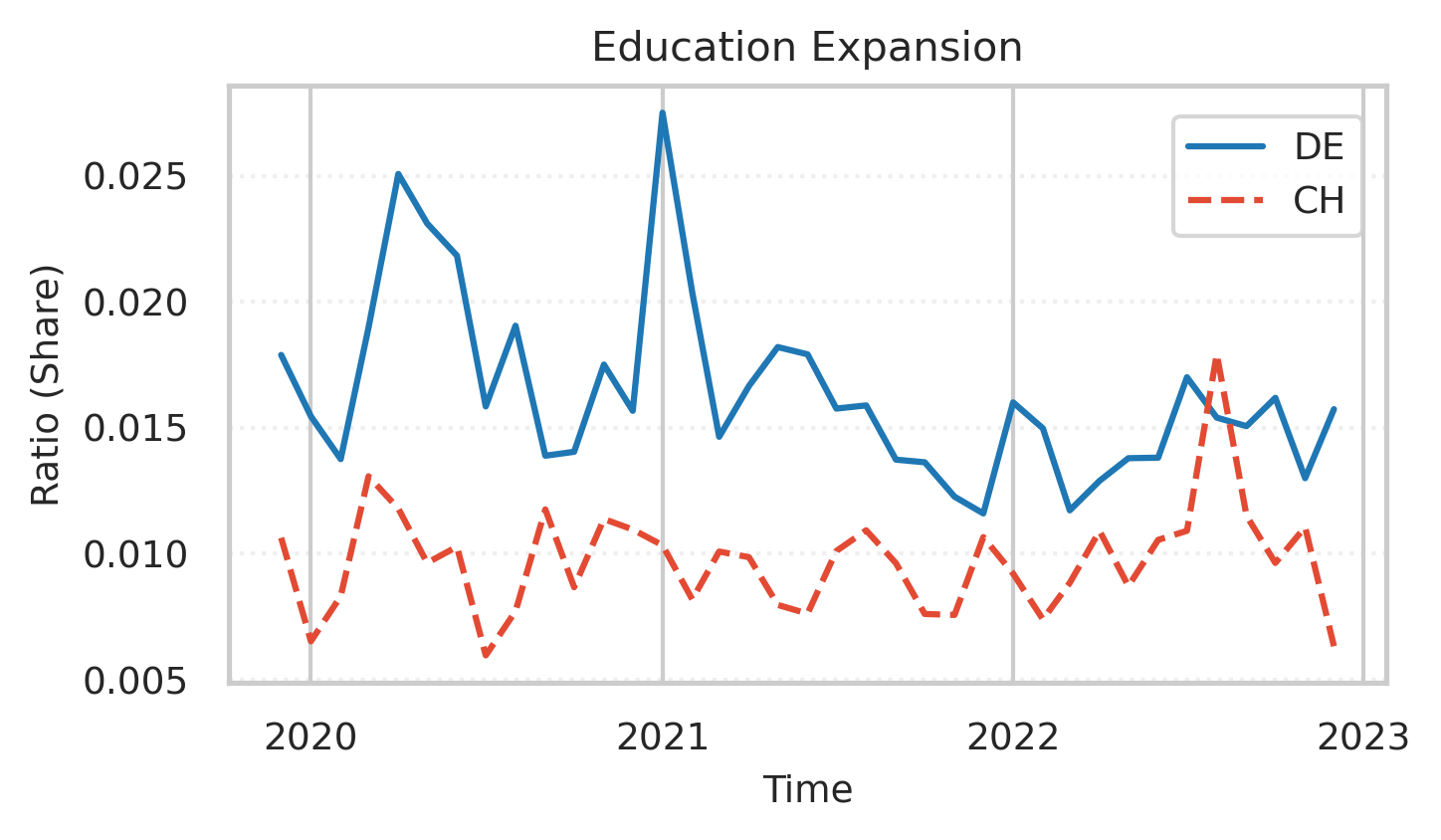}
        \label{fig:education}
    \end{subfigure}
    
    \caption{Monthly reporting ratios in German (DE) and Swiss (CH) news papers for the topics \textit{Federalism}($\rho = 0.34$) and \textit{Education Expansion} ($\rho = 0.09$) show the lowest correlations in the dataset. This could reflect national differences such as more direct participatory democracy in Switzerland and differences in COVID-19 measures in schools.}
    \label{fig:domain_trends_combined}
\end{figure}

As illustrated in Fig.~\ref{fig:domain_trends_combined}, these temporal dynamics could reflect the distinct federal and institutional structures of both countries: Throughout the COVID-19 pandemic, different countermeasures in different states of Germany led to incoherent policies across the nation, in contrast to Switzerland. The often as inconsistent perceived policies in different states have often led to heated debates in Germany on the topic of federalism. One of the most relevant political topics in the domain \textit{Welfare and Quality of Life} was education and the different policies for school lockdowns in Germany. Switzerland opted for a different policy on school lockdowns, which could be reflected in the low correlation of the political discourse across classical news media. 

\subsection{Temporal Dynamics in Swiss and German News}

Inspecting the overall temporal dynamics we observe that many political topics are driven by the global crisis events, however we also find distinct differences between political topics across nations. 
The \textit{Welfare State Expansion} ratio rises sharply in both countries following the WHO emergency declaration in January 2020, peaks with the first national lockdowns in March 2020, and remains elevated through 2021 before declining in spring 2022 to near pre-pandemic levels.

The \textit{Military: Negative} ratio remains low in both countries until early 2022, then surges simultaneously in February 2022, coinciding with the Russian invasion of Ukraine. This spike is very sharp, representing one of the most distinct event-driven shifts in the entire observation period.

The \textit{Democracy} category shows notable country-specific variation. In Switzerland, elevated values align with the COVID-19 law referenda in June and November 2021. In Germany, the peak corresponds to the federal election in September 2021. 

\begin{figure}[H]
    \centering
    \includegraphics[width=\textwidth]{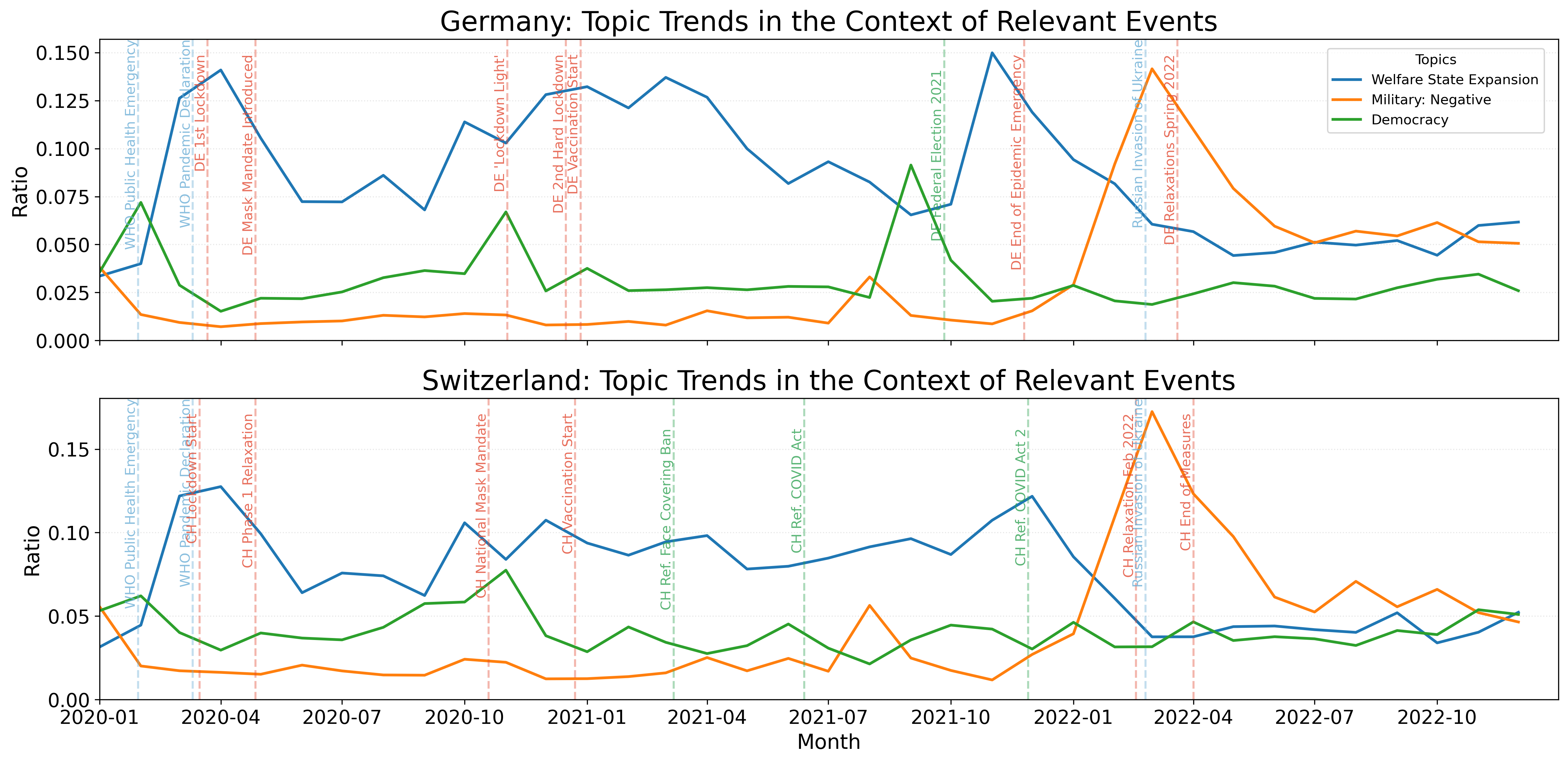}
    \caption{Comparative analysis of Welfare State Expansion, Military, and Democracy reporting in Germany and Switzerland (2020--2022). Both countries show closely aligned trends in \textit{Military: Negative} and \textit{Welfare State Expansion}, reflecting shared responses to global events. In contrast, \textit{Democracy} reveal pronounced national differences.}
    \label{fig:domain_trends}
\end{figure}

\subsection{Simplifying Political Preferences to Right-Left Orientation}
While the fine-grained political taxonomy of the Manifesto coding scheme allows for a detailed view into the political discourse, it can be difficult to relate these results to simpler schemes of political analyses, such as a binary political preference for two parties only. Next to the aggregation into seven political domains, the Manifesto coding scheme also offers an aggregation into right and left wing topics. This simpler dichotomy of political preferences allows for a more direct comparison of temporal dynamics with simpler ideological analyses.

Right-coded topics consistently account for 35\% of all classified content in both countries, while left-coded topics represent approximately 20\%. The two countries' time series run closely in parallel. A notable joint spike in left-coded content occurs in spring 2022, attributable primarily to the \textit{Military: Negative} category, which is mapped to the left dimension in the RILE scheme.

Switzerland exhibits a slightly but consistently higher share of right-coded content than Germany, with a gap of approximately 3--4 percentage points. This aligns with findings in the literature on greater emphasis on personal responsibility in Swiss COVID-19 coverage~\cite{zimmermann_newspaper_2023}, and may be partly explained by the Swiss sub-corpus being dominated by economically liberal outlets (NZZ, Tages-Anzeiger, SRF).

\begin{figure}[H]
    \centering
    \includegraphics[width=\textwidth]{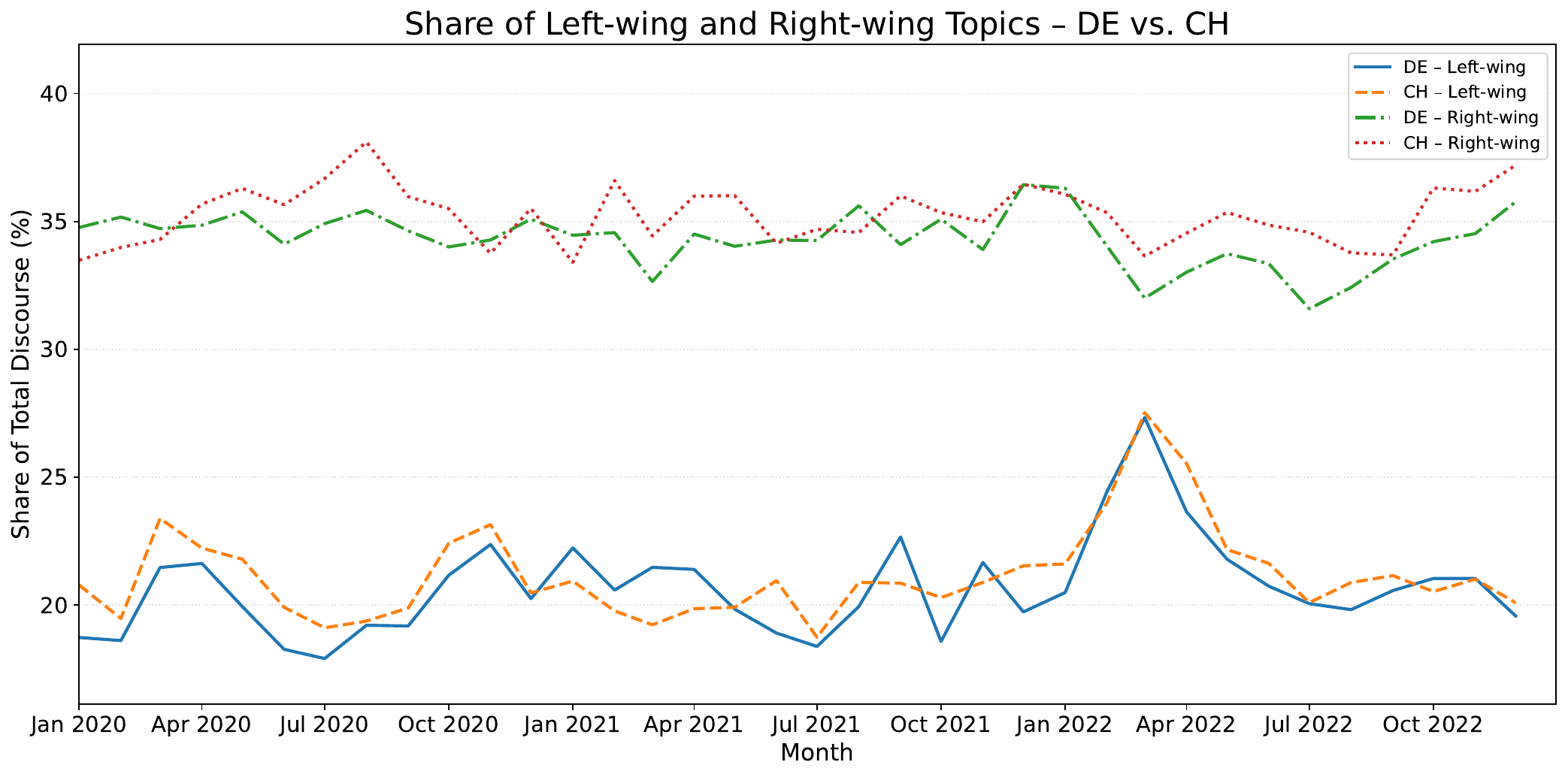} 
    \caption{Monthly share of left-wing and right-wing topics in the public german discourse, classified based on the Manifesto Project coding scheme. Across both countries, left-wing and right-wing shares remain relatively stable and balanced throughout the observation period. A notable exception occurs in early 2022, where a pronounced spike in left-wing topics is observable in both countries, coinciding with the Russian invasion of Ukraine.}
    \label{fig:ideological_trends.}
\end{figure}

\section{Discussion}
The following discussion analyses the results along three dimensions: media forms, national context and ideological orientation and reflects on what these patterns reveal about the role of global crises in shaping political discourse.
\subsection{Thematic Convergence and Divergence across Pandemic Phases}

The analysis of thematic domains illustrates how different media environments respond to a global crisis. 

In the pre-pandemic period, the newspaper agenda was stable and dominated by \textit{Economy} topics, reflecting an editorial logic in which fixed sections and thematic hierarchies produces a consistent agenda. Twitter, by contrast, has a fragmented distribution, with no single domain persistently exceeding 20 \%. Without a major external event, attention on Twitter remains dispersed, and no shared thematic focus emerges.

During the core pandemic phase (2020-2021), these differences temporarily narrow. The synchronous rise of  \textit{Welfare and Quality of Life} in both media indicates that the pandemic changes the previous media discourse, bringing their thematic focus closer together. Yet the intensity of Twitter's response to specific events far exceeds that of the newspaper corpus. The prominent spikes in March 2020 and spring 2021 reflect the event-driven nature of social-media discourse, where collective mobilization around immediate political concerns produces extreme short-term peaks, leaving no lasting imprint on the overall discourse. Newspapers handle the same events through established editorial routines, producing more gradual thematic adjustments.

After the core pandemic phase, the two media forms diverge again. Twitter experiences an extreme spike in \textit{Political System} around the German federal election before fragmenting rapidly, while the newspaper agenda remains comparatively stable. This suggests that while the pandemic initially forced a synchronization of topics, it did not lead to a lasting convergence of thematic focus between the media. Once the immediate reaction fades, attention on Twitter starts to shift to another topic, illustrating that social media attention is captured by events but not sustained beyond them. Newspapers, by contrast, provide a more balanced thematic distribution.


\subsection{Convergence Driven by International Events}
While the pandemic already proven to be a powerful synchronizing force across media forms, this effect appears even stronger across national borders.

The strong thematic similarity between German and Swiss media coverage confirms that international events generate a shared attention focus that temporarily overrides national differences in media structure and political culture. The near-identical temporal trajectories of \textit{Welfare State Expansion} and \textit{Military: Negative} across both countries illustrate how global crises synchronize national media agendas. This complements prior findings on cross-cultural COVID-19 reporting~\cite{sato_cross-cultural_2022}, showing that within the German-language space, commonalities dominate where cultural and linguistic distance is small. However, this convergence is not uniform across all topics.

\subsection{National Divergence in Domestically Anchored Topics}

In Switzerland, \textit{Federalism} shows a pronounced spike in early 2021, which may coincide with the temporary centralization of power during the national emergency, whereby the otherwise highly autonomous cantons ceded significant decision-making authority to the central government. Germany shows a similarly volatile pattern, though without a comparable peak, possibly reflecting the absence of legally binding national detail regulations, with individual states retaining greater autonomy throughout~\cite{desson_europes_2020}. 

The low correlation in \textit{Education Expansion} plausibly reflects these same asynchronous decision-making logics. Given that central pandemic measures in the DACH countries were implemented at staggered intervals despite similar institutional structures, it is plausible that education policy debates, did not unfold synchronously across the two countries. 


The \textit{Democracy} category reflects each country's distinctive political mechanisms. In Switzerland, elevated values tend to coincide with the COVID-19 law referenda in June and November 2021, as well as with specific national measures , reflecting the role of direct democracy in shaping pandemic governance. In Germany, the peak broadly corresponds to the federal election in September 2021 and to policy decisions such as national lockdowns. These contrasting peaks illustrate how the same topic is driven by fundamentally different political mechanisms. While Switzerland's direct democratic process gives citizens a direct vote on pandemic legislation, Germany's  democracy operates through different decision-making channels, producing distinct patterns in political discourse.

\subsection{Ideological Analysis: Interpretation and Limits}

The dominance of right-coded topics in both countries reflects the thematic composition of the RILE classification rather than an explicit editorial stance. Categories such as \textit{Freedom and Human Rights} and \textit{Constitutionalism} are classified as right-coded by the Manifesto scheme, and pandemic reporting naturally foregrounds these themes. The left--right metric captures \textit{topical frequency}, not the normative position taken in individual articles. The observed 3--4 percentage point difference between Switzerland and Germany should therefore be interpreted as a slight difference in thematic emphasis rather than as evidence of fundamentally divergent ideological orientations.

\section{Limitations}

Several limitations should be considered. First and foremost both text corpora suffer from sampling biases that are difficult to control for and correct. For the Twitter data, the original corpus~\cite{kratzke_monthly_2020} as well as the filtered version used in this study is subject to proprietary sampling protocols. Similarly we observe strong biases, for instance with respect to the number of articles in German and Swiss news outlets. The German sub-corpus (3.1 million articles) dwarfs the Swiss sub-corpus (ca.\ 98,000 articles); relative shares were used to mitigate this imbalance, but rare categories in the Swiss data remain subject to higher random variance. These sampling biases across nations and media types lead to heavily skewed distributions when grouping by sources: \textit{Die Welt} alone accounts for nearly 800,000 German articles, while the Swiss corpus is dominated by the NZZ, Tages-Anzeiger, and SRF. A robustness check removing \textit{Die Welt} produced qualitatively similar results, but source concentration remains a potential confound. Finally, the CC-NEWS archive contains gaps for individual months, particularly in the Swiss data, which may affect analyses of specific time windows. 

\section{Conclusion}

This study investigates the potential of automated political text classification for large-scale comparative media analysis. Our analysis includes several million tweets and news paper articles categorized in a fine-grained taxonomy of political preferences~\cite{lehmann_manifesto_2025} using ML models trained with a dedicated data set annotated by political experts. 

The results reveal a clear duality in the German-language media landscape. German and Swiss news coverage exhibit a high degree of thematic similarity in response to global events, with international crises acting as a powerful synchronizing force that temporarily overrides national differences. At the same time, domestically anchored topics continue to reflect each country's distinct institutional and political context, with divergences emerging precisely where national policies and political cultures play a defining role. The comparison between newspapers and Twitter further underscores that the medium itself shapes the rhythm of political discourse: while newspapers provide a stable editorial anchor, Twitter functions as an event-driven amplifier characterized by intense but short-lived spikes in attention.

Future work should extend this analysis to the full DACH region and investigate cross-linguistic comparisons. Furthermore, refining the classification model to better capture sentiment and framing, would enable a more nuanced assessment of political bias in online media, bringing automated political analysis closer to the transparency and accountability that informed public discourse requires.


\section*{Acknowledgements}
We thank the anonymous reviewers for their constructive feedback and Pola Lehmann at the Manifesto Project for her continuous support. This research was supported by the Einstein Center Digital Future, Berlin, and by the German Research Foundation (DFG) - Project number: 528483508 - FIP 12.

\printbibliography

\end{document}